# Exhuming nonnegative garrote from oblivion using suitable initial estimates- illustration in low and high-dimensional real data

Edwin Kipruto[1], Willi Sauerbrei[1]

[1] Institute of Medical Biometry and Statistics, Faculty of Medicine and Medical Center - University of Freiburg, Stefan-Meier-Street 26, 79104 Freiburg, Germany

**Summary**

The nonnegative garrote (NNG) is among the first approaches that combine variable selection and shrinkage of regression estimates. When more than the derivation of a predictor is of interest, NNG has some conceptual advantages over the popular lasso. Nevertheless, NNG has received little attention. The original NNG relies on least-squares (OLS) estimates, which are highly variable in data with a high degree of multicollinearity (HDM) and do not exist in high-dimensional data (HDD). This might be the reason that NNG is not used in such data. Alternative initial estimates have been proposed but hardly used in practice. Analyzing three structurally different data sets, we demonstrated that NNG can also be applied in HDM and HDD and compared its performance with the lasso, adaptive lasso, relaxed lasso, and best subset selection in terms of variables selected, regression estimates, and prediction. Replacing OLS by ridge initial estimates in HDM and lasso initial estimates in HDD helped NNG select simpler models than competing approaches without much increase in prediction errors. Simpler models are easier to interpret, an important issue for descriptive modelling. Based on the limited experience from three datasets, we assume that the NNG can be a suitable alternative to the lasso and its extensions. Neutral comparison simulation studies are needed to better understand the properties of variable selection methods, compare them and derive guidance for practice.

# 1. Introduction

Variable selection plays an important role in regression analysis, and several methods have been proposed depending on the goal of the analysis. Descriptive modeling is one of the most commonly used approaches that mainly builds on the data modeling culture. It aims at capturing the data structure parsimoniously rather than obtaining optimal predictive performance. Nevertheless, a suitable descriptive model can also be an acceptable predictive model. Descriptive models are often simpler than predictive models and are thus advantageous when interpretability, transportability, and general usability are important criteria to consider (Shmueli 2010; Sauerbrei et al. 2020). In this paper, we will consider both descriptive and predictive modeling.

Classical variable selection approaches have been widely used for decades. However, they have several drawbacks, the most severe being lack of stability; thus the resulting model has poor prediction in new data (Sauerbrei and Schumacher 1992; Breiman 1996). Best subset selection is computationally infeasible for a larger number of variables, and stepwise deletion is an efficient alternative to best subset selection but does not guarantee to find the best possible model (Miller 2002; James et al. 2013). Modern variable selection methods that allow for high-dimensional statistical inference have been proposed (Buehlmann and van de Geer 2013). Penalized methods are part of modern methods for variable selection that mitigate some of the computation and discrete problems of classical methods (i.e., covariates are either retained or dropped from the model). They combine variable selection and shrinkage, and are also continuous processes that shrink coefficients towards zero, reducing high variability which can improve prediction accuracy of models (Hastie, Tibshirani and Friedman 2009; Breiman 1996). Numerous studies have focused on penalized regression methods like the least absolute shrinkage and selection operator (lasso) (Tibshirani 1996),

Smoothly Clipped Absolute Deviation (SCAD) (Fan and Li 2001), and adaptive lasso (Zou 2006).

The nonnegative garrote (Breiman, 1995) and the lasso are among the first approaches that combine variable selection and shrinkage, but in practice, the former has received little attention despite some of its good conceptual properties. For instance, Zou (2006) showed that for a fixed number of variables the lasso is in general not variable selection consistent unless the design matrix satisfies strong assumptions, while the nonnegative garrote does not require such strong assumptions. Moreover, he showed that the lasso shrinkage produces biased estimates for the large coefficients because it shrinks both large and small nonzero coefficients equally, which can be detrimental to prediction due to excessive amount of bias. On the other hand, the NNG imposes severe shrinkage on small coefficients while large coefficients are hardly shrunken, which is a desirable property, especially when the regression estimates are of primary interest.

This paper focuses on the NNG that was originally proposed for modeling in classical linear regression models in low-dimensional data. The method has good features of both subset selection and ridge regression and is said to select simpler models with good predictive accuracy (Breiman 1995). The NNG like adaptive lasso requires initial estimators from the full model to be used as weights for penalizing different coefficients. It has been shown that the NNG with OLS initial estimators is consistent in variable selection in low-dimensional data when collinearity is not a concern, and the tuning parameter is properly chosen (Zou 2006; Yuan and Lin 2007).

A major drawback of the original NNG is its explicit reliance on OLS estimators, which perform poorly in highly correlated settings (Tibshirani 1996; Yuan and Lin 2007). Similarly, when the number of unknown parameters is much larger than the sample size, the OLS estimator is unavailable, and the NNG cannot be applied. About 15 years ago, it was shown that NNG can be used with other initial estimators. For instance, ridge initial estimators have been used in highly correlated settings because they are more stable than OLS estimators (Yuan and Lin 2007). In high-dimensional settings, Zhang, Jeng and Liu (2008) investigated the theoretical properties of ridge and lasso as initial estimators for two-step procedures, including the NNG. Their study showed that the ridge estimator can be used as an initial estimator when the tuning parameter is properly chosen. Despite many citations (Google scholar 1441 on October 27, 2022) of the original article by Breiman (1995) and the possibilities offered by proposed initial estimates, it seems that NNG is hardly used in practice. The lasso and some of its extensions are the dominating methods.

Penalized regression estimators including the NNG do not fit well into classical theory because the resulting estimators are biased (Van Houwelingen 2001). Several standard error estimators have been proposed for inference using penalized estimators. Fan and Li (2001) showed that the sandwich formula can be used as an approximate estimator for the covariance of the SCAD estimates. Xiong (2010) derived the sandwich formula for the NNG estimator. It is important to note that the sandwich formula gives an estimated variance of zero for covariates with zero coefficients, which is unsatisfactory. The bootstrap has been suggested as an alternative method for estimating the standard errors of the penalized estimators (Tibshirani 1996).

Using three publicly available datasets, this study sought to demonstrate that nonnegative garrote: (i) performs well in low-dimensional data, both with a low and high correlation in terms of variable selection and prediction, and (ii) can be applied in high-dimensional settings, which would imply that the NNG may be a suitable alternative to popular approaches. In low-dimensional data with low correlation, we will investigate the effects of using penalized and unpenalized initial estimators on the NNG regression coefficient estimates, model selection and prediction. In addition, we will compare the proposed sandwich and bootstrap standard errors of the NNG estimates. In all datasets, the performance of the NNG was compared with the lasso, adaptive lasso, relaxed lasso and best subset selection.

The rest of the paper is organized as follows. Section 2 describes the three data sets used and discusses the relevance of data standardization. Section 3 describes the NNG with its tuning parameters, initial estimators, and standard errors of estimates. Section 4 compares the results of NNG with those of competing approaches. Section 5 contains the discussion and conclusions. Due to space limitations, software implementation and a detailed description of other variable selection methods have been relegated to the supporting information.

## 2. Datasets

Three datasets with different structures were used, two of which were low-dimensional with a low and relatively higher degree of multicollinearity, while one was high-dimensional. The variance inflation factors (VIF) and conditional number (CN) of the design matrix were used to quantify collinearity in the design matrix (Belsley, Kuh and Welsch 2005). To demonstrate the performance of approaches in a low degree of collinearity, we reanalyzed the prostate cancer data set from the study by Stamey et al. (1989). The data consisted of the medical records of 97 male patients who were about to undergo radical prostatectomy. The response variable was the logarithm of prostate-specific antigen, while the covariates were eight clinical measures. The VIF ranged between 1.34 and 3.10, while the CN was 4.15; an indication of a low degree of collinearity.

In addition, we reanalyzed the body data by Johnson (1996) containing records of physical and body circumference measurements of 252 men. The outcome was the percentage of body fat and 13 covariates. Based on a detailed check for influential points (IP), two IPs (39 and 216) were eliminated before analysis. The CN was 21.06, while the VIF ranged from 1.82 to 45.32, indicating potential collinearity problems. We reanalyzed the preprocessed data set reported by Boulesteix, Guillemot and Sauerbrei (2011) to illustrate that fitting the NNG in high-dimensional data is feasible. This data was initially used to develop gene-expression prediction models for disease outcomes for (n=286) patients with lymph node-negative breast cancer. The outcome was a binary variable with 1 denoting a relapse ($n_1 = 107$) and 0 denoting no relapse ($n_2 = 179$), with a total of 22,283 covariates. All the datasets used are publicly available and more details are given in Web Appendix A of the supporting information.

Standardized covariates are generally recommended before fitting penalized regression methods unless they are all measured in the same units (James et al. 2013). However, unlike other methods, the NNG is scale-invariant (Breiman 1995). The standard deviation of each column of the design matrix was used to scale the original covariates in all datasets to ensure that all covariates were on the same scale. This approach is sensible when all covariates are continuous and assumed to be linearly related to the outcome variable. This assumption was acceptable in the present study since some continuous variables that were deemed to be nonlinear, for instance, in prostate cancer data, were logarithm transformed. The approach is problematic in the presence of binary variables because the high prevalence cells will dominate the penalty function in penalized methods (Harrell 2016). Nevertheless, we resolved to use this approach since we only had a single binary covariate in one dataset.

## 3. Methods for variable selection

In penalized likelihood procedures, we considered the NNG, lasso, adaptive lasso (Alasso), and relaxed lasso (Rlasso) (Meinshausen 2007), while the best subset selection was considered in classical variable selection strategies. In this section, the NNG is discussed in depth, while other methods are described in Web Appendix B of the supporting information.

**3.1 Nonnegative garrote**

The original NNG estimator consists of initial estimation of OLS estimates, $\hat{\beta}^{OLS}$, from the full least-squares model, the selection of the tuning parameter, λ, and the estimation of nonnegative shrinkage factors c=($c_1,…,c_p$ )$^T$ (Kipruto and Sauerbrei 2022). The same process can be applied to generalized linear models by replacing OLS estimates with maximum likelihood estimates. For classical linear regression models, the shrinkage factors $\hat{c}(\lambda)$ are obtained by optimizing

$$\hat{c}(\lambda) = arg\,min_c \frac{1}{2n}\sum_{i=1}^{n}\left(y_i - \sum_{j=1}^{p}c_j x_{ij}^*\right)^2 + \lambda\sum_{j=1}^{p}c_j, \qquad c_j \geq 0, \qquad \lambda \geq 0.$$

where $x_{ij}^* = \hat{\beta}_j^{OLS} x_{ij}$ and $\lambda$ is a tuning parameter. The NNG estimate is calculated as $\hat{\beta}_j^{NNG}(\lambda) = \hat{c}_j \hat{\beta}_j^{OLS}$. When the columns of $X$ are orthogonal, i.e., $X^T X = I_n$ the shrinkage factors can be estimated using $\hat{c}_j(\lambda) = \left(1 - \frac{\lambda}{\left(\hat{\beta}_j^{OLS}\right)^2}\right)_+ = m_+$, where $m_+ = \max(m, 0)$ denotes the positive part of *m*. This implies that the regression coefficients whose OLS estimates are large in absolute terms in a full model will have shrinkage factors close to 1, while noise covariates are likely to have OLS estimates close to zero and as a result, the shrinkage factors can be exactly zero (Yuan and Lin 2007; Kipruto and Sauerbrei 2022), as shown on the top-left panel of **Figure 1**. When $\lambda = 0$, the penalty term has no effect and all shrinkage factors are equal to 1 and the NNG estimates are equal to OLS estimates. On the other hand, when $\lambda \to \infty$, all shrinkage factors are equal to zero and the NNG gives the null model in which all regression estimates are equal to zero. This means that the performance of NNG critically depends on the tuning parameter. In estimating the tuning parameter for the NNG, Breiman (1995) found that 10-fold cross-validation (CV) was more reliable than leave-one-out cross-validation (LOOCV). For this reason, we estimated $\lambda$ using 10-fold CV, where the optimal $\lambda$ was obtained by selecting the model that minimized the mean squared error and deviance in Gaussian and binomial models, respectively.

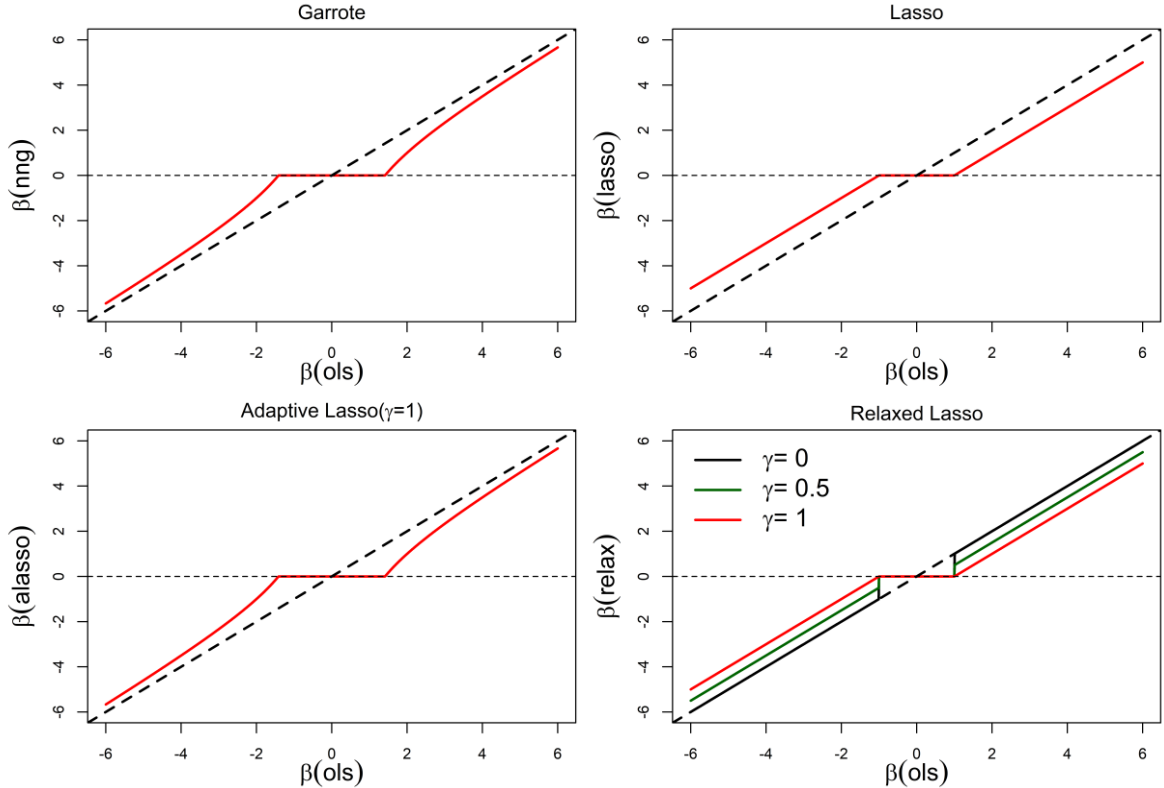

**Figure 1**. Shrinkage behavior of penalized methods in orthogonal design. The estimate of each procedure (y-axis) is plotted against the OLS estimate (x-axis). The dashed line is the line of equality. Adapted from several authors (Tibshirani 1996; Zou 2006; Meinshausen 2007).

### 3.1.1 Initial estimates for nonnegative garrote and adaptive lasso

The choice of initial estimates is crucial for the success of the NNG and Alasso. The original NNG relies on OLS estimates, which are known to be highly variable in a high degree of multicollinearity, and its unique solution does not exist in high-dimensional settings, thus both phenomena have a negative impact on the NNG. The latter could be the probable reason that NNG is not used in the analysis of high-dimensional data (Kipruto and Sauerbrei 2022). However, Yuan and Lin (2007) demonstrated that NNG is a flexible approach that can be used with other initial estimators such as the ridge or elastic net. Several initial estimators have been proposed for the Alasso. In low-dimensional data, Zou (2006) recommended the use of OLS estimators except when multicollinearity is problematic, in which case he proposed ridge estimators. In high-dimensional settings, Huang, Ma and Zhang (2008) proposed univariate regression estimators in situations where the zero and nonzero components are uncorrelated or at most weakly correlated, which is an unrealistic assumption in most applications. Furthermore, lasso initial estimators have been proposed. In this study, we used OLS, ridge, and lasso estimators as initial estimators for NNG and Alasso. The latter two were tuned in a prediction optimality way using 10-fold CV and were used for both low and high-dimensional data.

**3.1.2 Standard errors of NNG regression estimates**

It has been shown that local quadratic approximation (LQA) can provide a sandwich formula for variance estimation of nonzero components for the SCAD penalty (Fan and Li 2001). Through simulation studies, Fan and Li (2001) showed that the formula has good accuracy even for moderate sample sizes. In addition, Zou (2006) used the LQA approach to derive a sandwich formula for the adaptive lasso and tested its accuracy in simulation studies, again reporting that the standard error formula works quite well. The LQA approach was also used by Xiong (2010) to derive the standard error formula for the NNG estimator. The sandwich formula gives an estimated variance of zero for covariates with zero coefficients, which is unsatisfactory.

The bootstrap has been suggested as an alternative method (Tibshirani 1996) and several studies have been conducted to evaluate the performance of the bootstrap standard errors. For instance, Knight and Fu (2000) studied the asymptotic behavior of the lasso estimator using residual bootstrap and established that when there are one or more zero components, the bootstrap approximation may fail in consistency. They argued that one possible solution to this is to use consistent model selection procedures like the adaptive lasso. Based on the fact that the NNG is a consistent model selection procedure (Zou 2006; Yuan and Lin 2007), it suffices to use bootstrap to estimate the variance of the NNG estimator. We used the nonparametric bootstrap method (Efron and Tibshirani 1994), in which a bootstrap sample was drawn with replacement. A total of 1,000 repetitions were conducted. To study the behavior of the bootstrap standard errors, the NNG was fitted in each bootstrap sample by either fixing λ at its optimal (or one standard error (1SE) rule) value from the original data or re-estimating it in each bootstrap sample using a 10-fold CV as conducted by Tibshirani (1996) for the lasso.

### 3.2 Optimal versus one standard error rule tuning parameters

In practice, a cross-validation scheme is often used to select tuning parameter(s) with the aim of achieving prediction optimality. When the goal is to recover the true model or select a simple model for description, a tuning parameter with a larger value than that of optimal prediction is required (Buehlmann and van de Geer 2013). As a result, a 1SE rule has been proposed in which the simplest model, whose error is within one standard error of the minimum error, is selected (Hastie *et al.* 2009). This approach was previously applied by Breiman *et al.* (1984) in the context of regression trees, where they reported a reduction in the instability of trees and helped choose the simplest tree whose prediction error was comparable to the optimal prediction error. In the current study, we compared the performance of optimal and 1SE rule tuning parameters in terms of model selection and prediction accuracy. In addition, we investigated the effects of optimal and 1SE rule tuning parameters on the proposed standard errors of NNG estimates.

### 3.3 Prediction errors

When developing prediction models, it is important to evaluate their prediction accuracy on new data, which is often not available. Several approaches for evaluating prediction performance that use the data at hand, such as data-splitting and cross-validation have been proposed (Hastie et al. 2009). The former requires sufficient data to allow for splitting of the data into a training set for model development and a test set for estimating test error. Since data are often scarce, this approach is not ideal because it reduces the sample size for model development and testing and may tend to overestimate the test error for the model fitted to the entire dataset (James et al. 2013; Harrell 2016). The approach also suffers from numerous other weaknesses as summarized by Harrel (2016). Cross-validation is an approach for estimating prediction error that uses all the data and seems to be widely accepted. In this study, 10-fold CV was used as recommended in practice due to a good compromise between bias and variance of the prediction error (Hastie et al. 2009; James et al. 2013). Two metrics were used as loss functions: (i) for Gaussian responses, we used the mean squared error (MSE), while (ii) for binary responses, we used the area under the receiver operating characteristic (ROC) curve, which is identical to the concordance statistic (Steyerberg 2020).

### 3.4 Notations

We introduce the notations used in the results section. The notation used here is somewhat different from the notation used in other studies but allowed us to efficiently deal with model comparisons. The NNG with OLS, ridge, and lasso as initial estimators in conjunction with cross-validation for selecting optimal tuning parameters is denoted by NNG (O, CVopt), NNG (R, CVopt), and NNG (L, CVopt), respectively. The Alasso with OLS, ridge, and lasso as initial estimators in conjunction with optimal tuning parameters was denoted by Alasso (O, CVopt), Alasso (R, CVopt), and Alasso (L, CVopt), respectively. When the 1SE rule was used to select tuning parameters, "opt" was replaced by "1se". Best subset selection with cross-validation and the BIC criterion was denoted by BS (CV) and BS (BIC) respectively.

## 4. Results

### 4.1 Prostate cancer data

**Nonnegative garrote initial estimates**

The choice of initial estimates is essential for NNG to correctly identify the set of relevant and irrelevant variables. In the prostate cancer study, it was sufficient to use OLS initial estimates from the full model due to the low degree of collinearity. However, we further investigated the performance of ridge and lasso initial estimates using optimal tuning parameters as shown in **Table 1** and compared the results. We also used the lasso estimates with a tuning parameter selected from the 1SE rule.

**Table 1.** Prostate data. Estimates of standardized covariates from four approaches used as initial estimates for NNG.

| Covariate | OLS | Ridge($\lambda_{opt}$) | Lasso($\lambda_{opt}$) | Lasso($\lambda_{ise}$) |
|---|---|---|---|---|
| x1 | **0.662** | **0.577** | **0.647** | **0.517** |
| x2 | **0.265** | **0.257** | **0.260** | **0.104** |
| x3 | -0.157 | -0.124 | -0.143 | 0.000 |
| x4 | 0.140 | 0.124 | 0.132 | 0.000 |
| x5 | **0.314** | **0.282** | **0.299** | **0.126** |
| x6 | -0.148 | -0.055 | -0.113 | 0.000 |
| x7 | 0.035 | 0.046 | 0.030 | 0.000 |
| x8 | 0.125 | 0.096 | 0.112 | 0.000 |
| #Variables | 8 | 8 | 8 | 3 |
| $R^2$ | 0.663 | 0.659 | 0.663 | **0.562** |
| Adj. $R^2$ | 0.633 | 0.628 | 0.632 | **0.548** |

The OLS and ridge estimates are slightly different, as shown in **Table 1** and **Figure 2** (left panel). Depending on the size of the tuning parameter, the lasso can force some of the estimates to be exactly zero. However, the optimal tuning parameter estimated via cross-validation did not eliminate any variable, and the resulting estimates were close to OLS estimates, indicating that the penalty term had minimal effects. Using the tuning parameter from the 1SE rule led to a model with only three nonzero coefficients, but the estimates were over-shrunken as compared to the OLS estimates (**Figure 2**, left panel) due to the larger tuning parameter; hence the model fit was affected, as shown by a smaller adjusted $R^2$ of 0.548 (**Table 1**). The two versions of the lasso estimates were used as initial estimates for NNG to show that NNG can be used to reduce the number of variables selected by the lasso and to correct for the over-shrinkage behavior of nonzero coefficients by the lasso. The former was achieved using the optimal lasso initial estimates and examining whether NNG further eliminated any variables; the latter involved a comparison of the Lasso ($\lambda_{1se}$) and NNG ($L(\lambda_{1se})$, $CV_{opt}$) estimates.

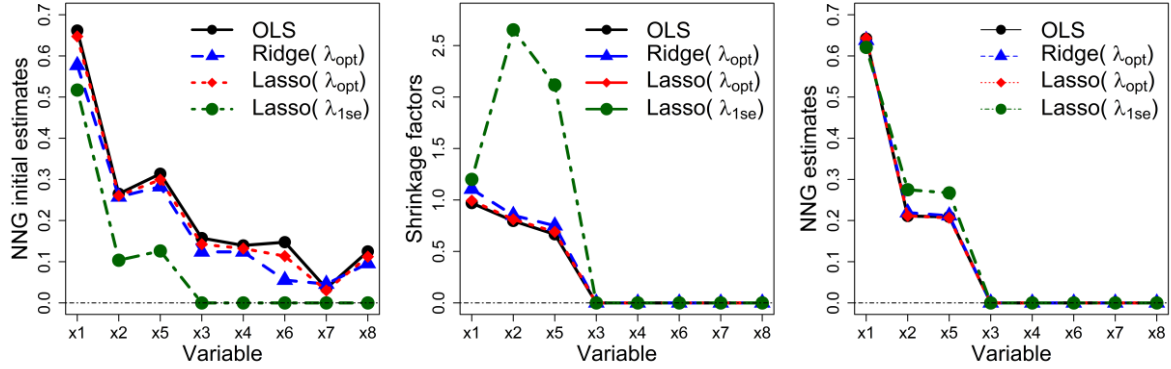

**Figure 2**. Prostate data. Left: the plot of absolute values of the NNG initial estimates. Middle: NNG shrinkage factors for different initial estimates. Right: NNG regression coefficients for different initial estimates

**Effects of initial estimates and tuning parameters on the NNG model selection**

**Table 2** shows the parameter estimates from the full OLS model and the number of variables selected by NNG using four different initial estimates (OLS, ridge ($\lambda_{opt}$), lasso ($\lambda_{opt}$), and lasso ($\lambda_{1se}$)). The results showed that the OLS model with eight variables explained about 66% of the total variation, and three variables (x1, x2 and x5) were significant at the 5% level, with the former having the strongest effects. Using any of the four initial estimates in conjunction with optimal tuning parameters from a 10-fold CV resulted in the selection of the three variables that were significant in the full model. A large difference in $R^2$ was observed between the lasso ($\lambda_{1se}$) ($R^2 = 0.56$, **Table 1**) and NNG with lasso ($\lambda_{1se}$) as initial estimates ($R^2 = 0.64$, **Table 2**). This indicates that the NNG can improve the model fit of the lasso by a larger margin by reducing the over-shrinkage of regression coefficients.

**Table 2.** Prostate data. OLS and NNG models. Effects of tuning parameters (second row) and initial estimates (third row) on the selected NNG model.

|  | Full OLS | | Nonnegative garrote | | | |
| --- | --- | --- | --- | --- | --- | --- |
|  | Estimate | p-value | OLS | ridge($\lambda_{opt}$) | lasso($\lambda_{opt}$) | lasso($\lambda_{1se}$) |
| x1 | **0.662** | **0.000** | x | x | x | x |
| x2 | **0.265** | **0.003** | x | x | x | x |
| x3 | -0.157 | 0.058 | - | - | - | - |
| x4 | 0.140 | 0.098 | - | - | - | - |
| x5 | **0.314** | **0.002** | x | x | x | x |
| x6 | -0.148 | 0.241 | - | - | - | - |
| x7 | 0.035 | 0.752 | - | - | - | - |
| x8 | 0.125 | 0.310 | - | - | - | - |
| #variable |  | 8 | 3 | 3 | 3 | 3 |
| $R^2$ |  | 0.66 | 0.63 | 0.63 | 0.63 | 0.64 |
| *Adj. $R^2$* |  | 0.63 | 0.62 | 0.62 | 0.62 | 0.62 |

"x" and "-" denotes that a variable is selected and eliminated, respectively

**Effects of initial estimates on NNG shrinkage factors and regression estimates**

**Figure 2** displays the absolute values of standardized initial estimates, the NNG shrinkage factors, and the NNG regression estimates. The values of the three initial estimates (OLS, ridge ($\lambda_{opt}$), and lasso ($\lambda_{opt}$)) were slightly different (left panel). As a result, their shrinkage factors (middle panel) and corresponding regression estimates (right panel) were almost identical. Two of the lasso ($\lambda_{1se}$) initial estimates (x2 and x5) that were nonzero are extremely small (left panel) and NNG recognized this and estimated shrinkage factors that were greater than one to correct for the over-shrinkage. In the full OLS model, the estimates for x1, x2 and x5 were 1.3, 2.6, and 2.5 times the estimates of the lasso ($\lambda_{1se}$), respectively. These values were close to the NNG shrinkage factors of 1.2, 2.7 and 2.1 for the aforementioned variables, respectively. This means that the resulting NNG regression coefficients were closer to the OLS estimates from the full model than the lasso ($\lambda_{1se}$), hence improving the model fit. In this case, the lasso acted as a screening method, and since its tuning parameter was large as a result of adopting the 1SE rule, no variable was further eliminated by the NNG, and instead a gain was observed in terms of reducing over-shrinkage of the lasso estimates. The right panel of **Figure 2** shows that NNG regression estimates are similar for all initial estimates.

**Nonnegative garrote standard errors**

The standard errors from the sandwich formula and the bootstrap were compared. Overall, the sandwich standard errors were relatively smaller than the bootstrap standard errors (Figure 3a and 3d) regardless of whether the optimal or 1SE rule tuning parameters were used. The two versions of bootstrap standard errors were in close proximity for variables with nonzero coefficients (x1, x2, and x5) in the original data, while large differences were observed for the zero coefficients (x3, x4, x6-x8), especially when optimal $\lambda$ was re-estimated in each bootstrap sample. This is because the covariates whose estimated coefficient was zero in the original data (x3, x4, x6-x8) were given nonzero coefficients in large proportions when $\lambda$ was re-estimated (Figure 3c) rather than fixed (Figure 3b). Adopting the 1SE rule decreased these proportions considerably (Figure 3e and 3f) as it selected the simplest model with strong covariates, though the bootstrap estimates were somewhat shrunken towards zero, especially for small nonzero coefficients (**Web Table 1** in the supporting information). Based on these findings, it seems that the sandwich standard errors for strong effects such as x1 are a good approximation to the bootstrap with fixed and re-estimated optimal $\lambda$, but poor approximations for moderate effects like x2 and x5.

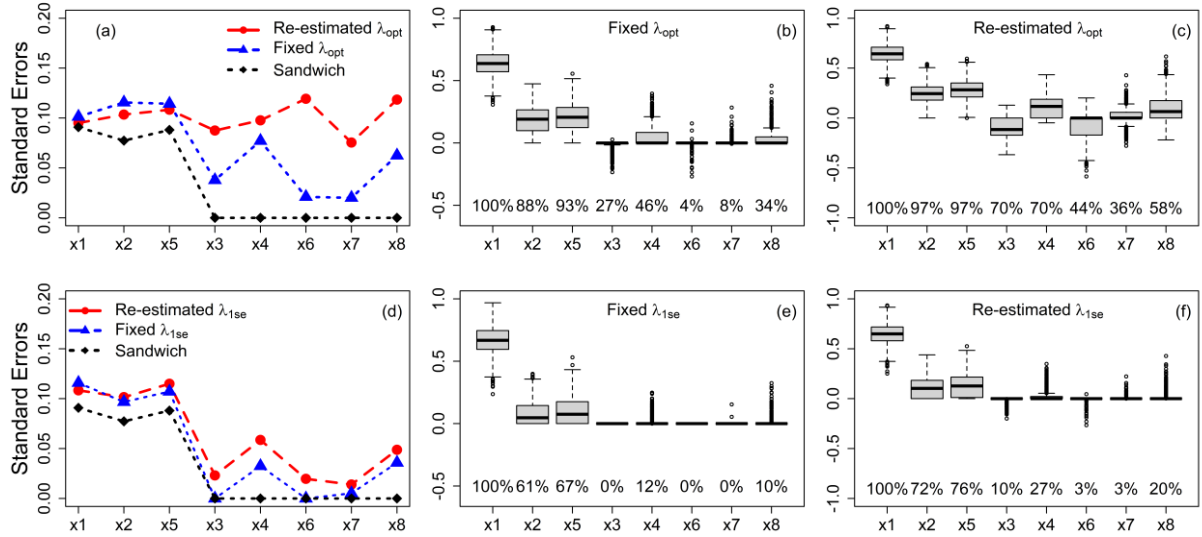

**Figure 3**. Prostate data. NNG (O, CV) standard errors (SE) and distribution of bootstrap estimates. Upper and lower panel: results using optimal and 1SE rule λ, respectively. Left panel (a and d): comparison of sandwich and bootstrap SE for re-estimated and fixed λ. Middle panel: distribution of bootstrap estimates for a fixed λ from the original data (b and e). Right panel: distribution of bootstrap estimates by re-estimating λ (c and f) in each bootstrap sample.

**Comparison of NNG with other selection procedures**

Table 3 shows the variables selected by different selection methods. The NNG and Alasso selected three covariates (x1, x2 and x5) regardless of the chosen initial estimates, and their model fits were nearly identical ($R^2$ of about 0.63). The same variables were also selected by relaxed lasso, best subset selection and lasso using the 1SE rule. However, no variable was eliminated by the lasso with the optimal tuning parameter. Even though the best subset selection using the CV and BIC criterion selected the same model, this is not always the case because the BIC usually selects smaller models than the CV because it penalizes models with many variables more heavily. The estimates of the relaxed lasso were identical to the estimates of the best subset selection without post-estimation shrinkage since its relaxation parameter ϕ =0, hence the model fit was identical. Generally, all selection methods fitted the data equally well except the lasso ($\lambda_{1se}$) model which performed poorly.

**Table 3.** Prostate data. Comparisons of variables selected and model fit for different selection methods applied. NNG and Alasso with OLS, ridge and lasso initial estimates selected same models.

| Variable | $\hat{\beta}^{OLS}$ | P-value | NNG | Alasso | Lasso($\lambda_{opt}$) | Lasso($\lambda_{1se}$) | Rlasso($\phi = 0$) | BSS(CV) | BSS(BIC) |
|---|---|---|---|---|---|---|---|---|---|
| x1 | 0.662 | **0.000** | x | x | x | x | x | x | x |
| x2 | 0.265 | **0.003** | x | x | x | x | x | x | x |
| x3 | -0.157 | 0.058 | - | - | x | - | - | - | - |
| x4 | 0.140 | 0.098 | - | - | x | - | - | - | - |
| x5 | 0.314 | **0.002** | x | x | x | x | x | x | x |
| x6 | -0.148 | 0.241 | - | - | x | - | - | - | - |
| x7 | 0.035 | 0.752 | - | - | x | - | - | - | - |
| x8 | 0.125 | 0.310 | - | - | x | - | - | - | - |
| #variables | 8 | | 3 | 3 | 8 | 3 | 3 | 3 | 3 |
| $R^2$ | | 0.66 | 0.63 | 0.63 | 0.66 | **0.56** | 0.64 | 0.64 | 0.64 |
| $Adj.R^2$ | | 0.63 | 0.62 | 0.62 | 0.63 | **0.55** | **0.63** | 0.62 | 0.62 |

"x' and "-" denotes variables selected and omitted respectively.

### 4.2 Body fat-highly correlated data

In this section, we assess the performance of NNG in highly correlated data and compare the results with other approaches. Our ultimate goal is to investigate whether replacing OLS initial estimators with ridge estimators can help mitigate the problem of multicollinearity and thus improve selected models and prediction performance.

**Comparison of variable selection methods in highly correlated settings**

Results of different procedures are summarized in Table 4. Variable x1 was found to be an important covariate of body fat since its removal led to a reduction in $R^2$ by about 14%. The relaxed lasso and BS (BIC) selected simple models with four and three variables, respectively, while the BS (CV) selected a complex model with 11 variables. Elimination of variables hardly influenced adjusted $R^2$, which ranged from 0.72 to 0.73.

The NNG (O, CVopt) selected a relatively larger number of variables (10 variables) than the NNG (R, CVopt) and NNG (L, CVopt) that selected eight variables each. The NNG (R, CVopt) and NNG (L, CVopt) selected the same variables even though the former started with all 13 variables, while the latter started with eight variables after five were removed in the first stage by the lasso. The NNG (L, CVopt) did not further remove any variables but reduced the shrinkage effect of the lasso (**Web Table 3** in the supporting information).

The Alasso (O, CVopt) and Alasso (R, CVopt) selected the same ten variables as NNG (O,

CVopt), implying that replacing OLS by ridge initial estimators had no effect on the model selected.

**Table 4.** Body fat data. Comparisons of variables selected and model fit for different selection methods. OLS, ridge and lasso initial estimates for NNG and Alasso are denoted by O, R and L respectively. Initial estimates are not applicable (NA) for lasso, relaxed lasso and best subset selection

|  | Full OLS | | | NNG | | | Alasso | | | Lasso | Rlasso | BS(CV) | BS(BIC) |
|---|---|---|---|---|---|---|---|---|---|---|---|---|---|
| Variable | Est. | P | %$R^2$red | O | L | R | O | L | R | NA | NA | NA | NA |
| Abdomen (x1) | 8.95 | 0.00 | 14.04 | x | x | x | x | x | x | x | x | x | x |
| Wrist (x2) | -1.65 | 0.00 | 1.71 | x | x | x | x | x | x | x | x | x | x |
| Age (x3) | 0.97 | 0.02 | 0.84 | x | x | x | x | x | x | x | x | x | - |
| Neck (x4) | -0.92 | 0.09 | 0.44 | x | x | x | x | x | x | x | - | x | - |
| Forearm (x9) | 0.58 | 0.16 | 0.29 | x | x | x | x | x | x | x | - | x | - |
| Thigh (x7) | 1.02 | 0.17 | 0.28 | x | x | x | x | x | x | x | - | x | - |
| Hip (x8) | -1.05 | 0.26 | 0.19 | x | - | - | x | x | - | - | - | x | - |
| Height (x5) | -0.55 | 0.27 | 0.18 | x | x | x | x | x | x | x | x | x | - |
| Biceps (x10) | 0.52 | 0.30 | 0.16 | x | x | x | x | x | x | x | - | x | - |
| Chest (x6) | -0.88 | 0.31 | 0.15 | x | - | - | x | x | - | - | - | x | - |
| Ankle (x13) | 0.28 | 0.43 | 0.09 | - | - | - | - | - | - | - | - | x | - |
| Weight (x11) | -0.68 | 0.71 | 0.02 | - | - | - | - | - | - | - | - | - | x |
| Knee (x12) | 0.02 | 0.97 | 0.00 | - | - | - | - | - | - | - | - | - | - |
| # variables | | | 13 | 10 | 8 | 8 | 10 | 10 | 8 | 8 | 4 | 11 | 3 |
| $R^2$ | | | 0.74 | 0.74 | 0.74 | 0.74 | 0.74 | 0.74 | 0.74 | 0.73 | 0.73 | 0.74 | 0.72 |
| Adj. $R^2$ | | | 0.73 | 0.73 | 0.73 | 0.73 | 0.73 | 0.73 | 0.73 | 0.73 | 0.72 | 0.73 | 0.72 |

**Comparison of fitted values of NNG with other selection procedures**

In Table 4, we established that some of the methods differed with respect to the covariates selected but fitted the data equally well with minor differences in explained variation. However, it is most unlikely that the fitted values from all observations will agree. Here, we investigated whether there was a reasonable agreement between the fitted values of NNG and other selection procedures using the Bland-Altman plot (Bland and Altman 1986). The plot compares fitted values of two methods by plotting individual differences (d) against the individual means. We summarized the lack of agreement by calculating the mean difference ($\bar{d}$) and the standard deviation of the differences (SD). We would expect most of the differences to lie within the limits of agreement (LOA) ) ($\bar{d} \pm 2SD$) in order to declare a satisfactory agreement (Bland and Altman 1986; Gerke 2020).

**Figure 4** shows the scatter plot and the Bland-Altman plot for four pairs of predictors. The fitted values of NNG (R, CVopt) and NNG (L, CVopt) are nearly identical (**Figure 4**a), with a mean difference (95% confidence interval) of 0.00 (-0.01, 0.01). The majority of the data points lie within the narrow LOA (-0.1, 0.1) (**Figure 4**b). In this case, the lasso or ridge initial estimators can be used interchangeably without much difference in model fit and fitted values. Based on these results, we only compared fitted values of NNG (R, CVopt) with other variable selection approaches. **Figure 4**c shows a linear relationship between the fitted values of NNG (R, CVopt) and the BS (BIC), with the difference ranging from -3.0 to 2.4 and a mean difference of 0.00 (-0.13, 0.13) (**Figure 4**d). This suggests that the fitted values of the two approaches were nearly identical on average, but the LOA (-2.12, 2.12) was much wider. A positive trend was evident in **Figure 4**f comparing the fitted values of NNG (R, CVopt) and the lasso. The scatterplot revealed that the fitted values of NNG (R, CVopt) tended to be smaller than those of the lasso on the lower end and vice versa on the upper end (**Figure 4**e). Generally, the level of agreement between NNG (R, CVopt), and different approaches varied, with evidence of good agreement with Alasso (L, CVopt) (**Figure 4**h) and poor agreement with BS (BIC).

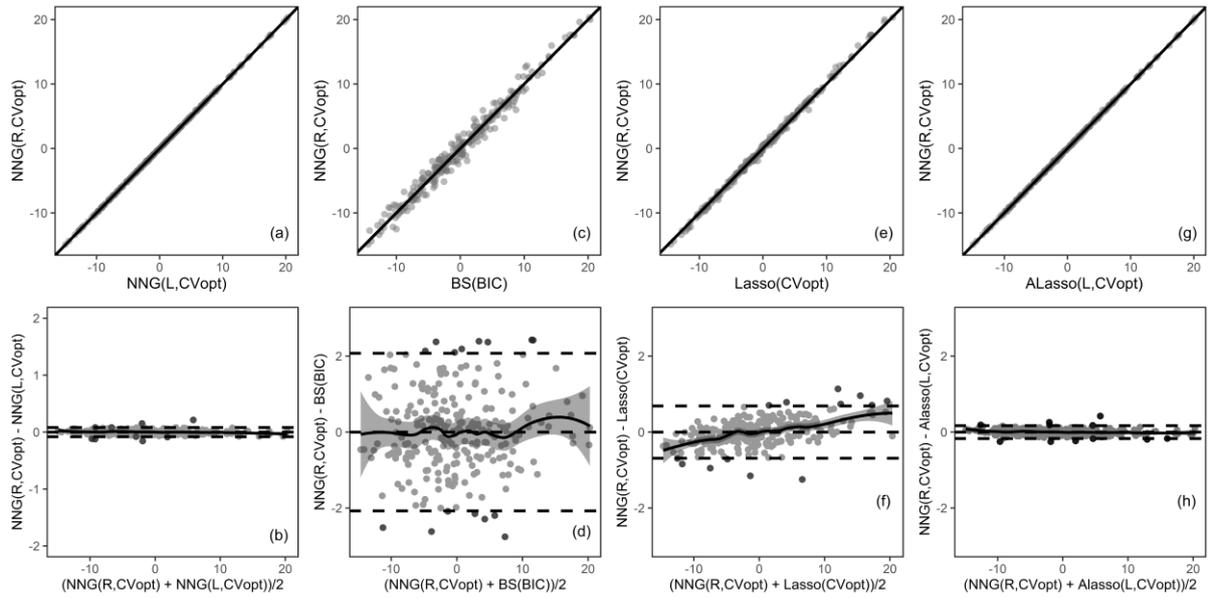

**Figure 4.** Body fat data. Fitted values from NNG, lasso, best subset selection, and adaptive lasso models. Upper panel: scatterplot with a line of identity. Lower panel: Bland-Altman plot for the difference against the mean of the two fitted values, with a 95% pointwise confidence interval (shaded and almost invisible). The horizontal dashed lines are drawn at the mean difference (always very close to zero) and the lower and upper LOA.

**Comparison of prediction errors in highly correlated data**

Within the NNG family of models, the NNG (R, CVopt) had a smaller MSE (**Figure** 5a), probably due to the stability of ridge initial estimates in highly correlated settings. A comparison of all the approaches showed that NNG (R, CVopt), Alasso (R, CVopt), and the lasso had slightly better predictive accuracy than the full OLS model, while BS (CV) performed the worst with high variation (**Figure** 5a and 5b). We also investigated the effects of choosing the tuning parameters of penalized methods using the 1SE rule on prediction errors (**Figure** 5c), and established that the percentage increase in prediction errors ranged from one (relaxed lasso) to six (adaptive lasso). Again, using the 1SE rule resulted in smaller models at the expense of prediction inaccuracy (**Figure** 5d).

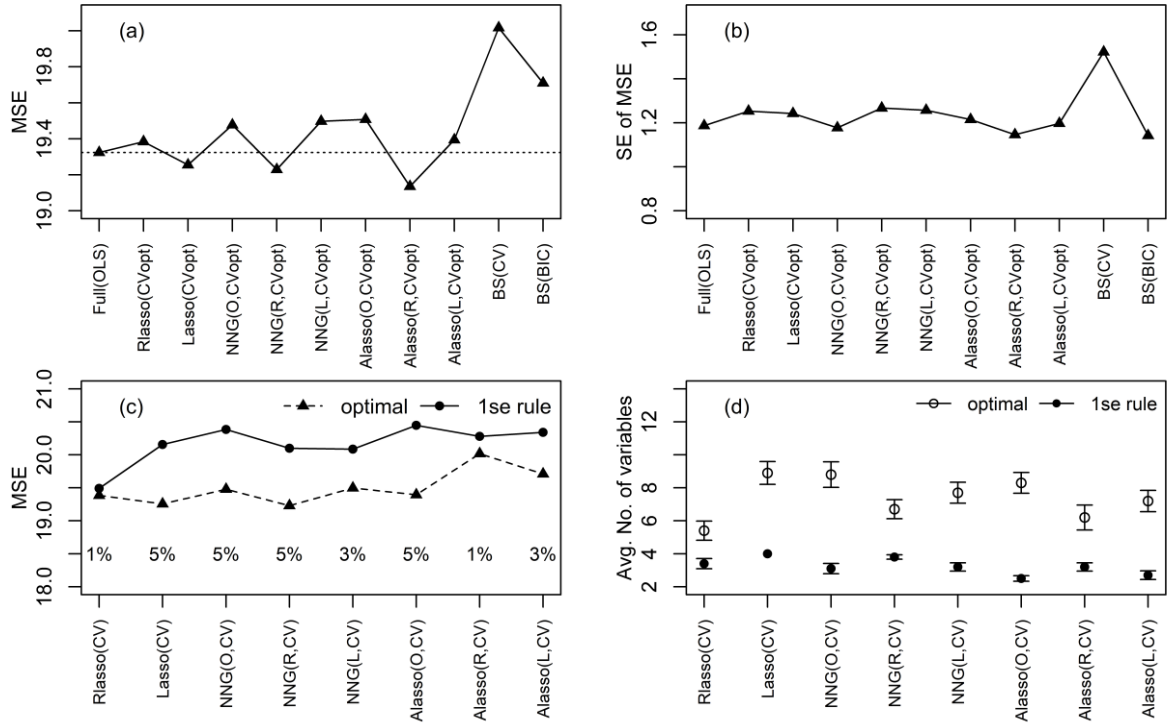

**Figure 5**. Body fat data. Cross-validation MSE for different methods, the dashed horizontal line is drawn at the MSE for the full OLS model. Top right: standard errors of cross-validation MSE. Bottom left: comparison of optimal and 1SE rule prediction errors of penalized methods. Bottom right: the average number of variables ((± one standard error bands) selected by penalized methods in CV using optimal and 1SE rule tuning parameters.

### 4.3 Gene expression - high dimensional data

In this section, we illustrate that NNG can be applied in high-dimensional data with a binary response variable. We compared the similarities and differences in variables selected using ridge and lasso initial estimates.

**Variable selection in high dimensional data**

**Table 5** shows dramatic differences in the number of variables selected by different approaches. When optimal tuning parameters were used, the NNG (L, CVopt) yielded a substantially sparser fit with 46 out of 22,283 variables, followed closely by Alasso (L, CVopt) with 57 variables. When ridge initial estimators were used, the NNG and Alasso selected larger models than the lasso and relaxed lasso. The lasso and relaxed lasso selected the same number of variables (94 out of 22,283) since the chosen relaxation parameter of the latter was ϕ=1; in this case, the two models were identical.

When the tuning parameters from the 1SE rule were used, the relaxed lasso and lasso selected an extremely small number of variables (one and two variables, respectively), compared to the NNG and adaptive lasso, using both ridge and lasso initial estimates. Further analysis of the variables selected by NNG and adaptive lasso using the ridge and lasso initial estimates revealed that they shared 40 and 36 variables when optimal and 1SE rule tuning parameters were used respectively (**Figure 6**).

**Table 5**. Gene expression data. Number of variables selected and AUC for different approaches using optimal and 1SE rule tuning parameters.

|  | Ridge | Lasso | Rlasso | Alasso(R) | Alasso(L) | NNG(R) | NNG(L) |
|---|---|---|---|---|---|---|---|
| # Variables (optimal) | 22 283 | 94 | 94 | 114 | 57 | 134 | **46** |
| # Variables (1SE) | 22 283 | 2 | 1 | 99 | 55 | 109 | 40 |
| AUC (optimal) | 0.709 | 0.662 | 0.626 | 0.673 | 0.671 | 0.688 | 0.658 |
| AUC (1SE) | 0.677 | 0.582* | 0.567* | 0.672 | 0.666 | 0.670 | 0.657 |
| % decrease in AUC | 4.5 | 12.1 | 9.4 | 0.1 | 0.7 | 2.6 | 0.0 |

* No variable was selected in some of the 10-folds, caused by large tuning parameters.

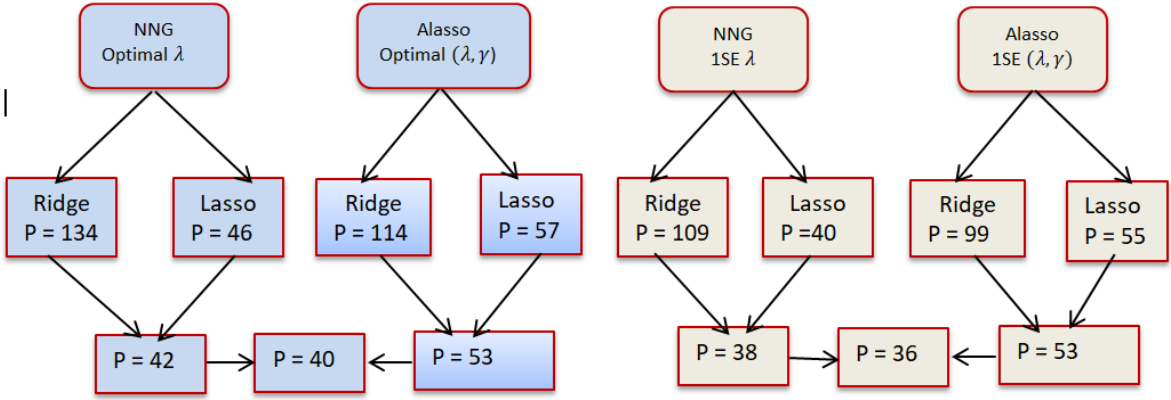

**Figure 6**. Gene expression data. The number of variables selected by nonnegative garrote and adaptive lasso using ridge and lasso initial estimators with optimal and 1SE rule tuning parameters. The approaches selected 40 and 36 variables in common when optimal and 1SE rule was used respectively. Further analysis of these 40 and 36 variables showed that 34 variables were in common.

**Prediction accuracy in high dimensional data**

**Table 5** shows the cross-validated AUC and **Web Table 5** (in supporting information) shows the average number of variables selected by different approaches in the CV. For optimal tuning parameters, all models had an AUC greater than 0.6 (**Table 5**) and they differed in accuracy. The ridge regression model had the highest AUC (0.709), while the relaxed lasso performed worst (AUC of 0.626). The performance of NNG (R, CVopt) and Alasso (R, CVopt) were very similar and slightly better than the lasso. In addition, the AUC of NNG (L, CVopt) and Alasso (L, CVopt) were similar to the lasso. However, the number of covariates selected in CV by the lasso was on average 69 variables, compared to 37 by the NNG (L, CVopt) and 45 by the Alasso (L, CVopt) (**Web Table 5** in supporting information).

Using the tuning parameters from the 1SE rule had a drastic impact on the prediction performance of lasso and relaxed lasso, while effects on NNG and adaptive lasso were negligible (**Table 5**). The poor performance of the lasso can be explained by the fact that the tuning parameter was too large such that all variables were removed from the model in two out of ten folds, and perhaps due to the effects of over-shrinkage of nonzero coefficients. For the relaxed lasso, it was possibly caused by a large tuning parameter, which led to underfitting because one variable was only selected in 4 out of 10 folds and no variable was selected in three of the 10 folds. Generally, the NNG with lasso initial estimates and the relaxed lasso selected smaller models on average in CV when optimal and 1SE rule tuning parameters were used, respectively (**Web Table 5** in the supporting information).

## 5      Discussion and conclusion

With the aim of exhuming NNG from oblivion in practice, we have assessed the performance of NNG in three real datasets with severe differences in the data structure. Our focus was on the variation in performance by comparing the effects of various initial estimates and tuning parameters in low and high correlation settings as well as high-dimensional data. Our results suggest that the NNG has some advantages, and at least it may be a worthy competitor to the other popular approaches. An R package for implementing NNG is in preparation.

Results from low-dimensional data with a low degree of collinearity showed that the NNG with optimal tuning parameters selected the same variables when various initial estimates were employed. Simulation experiments (Yuan and Lin 2007) have demonstrated that this is not always the case. The effects of using shrunken initial estimates were investigated, and we found that the NNG can correct for over-shrinkage of the initial estimates, thus improving the model fit. Yuan and Lin (2007) found similar results in a simulation study in which the NNG improved the estimation accuracy of an initial estimate. All approaches selected the same variables except for the lasso, which selected more variables. This is because the lasso often selects large models when optimal tuning parameters are determined using cross-validation (Buehlmann and van de Geer 2013; Meinshausen 2007).

In highly correlated data, the NNG with ridge initial estimates outperformed the NNG with OLS initial estimates by selecting smaller models with better prediction accuracy. This was probably due to the stability of the ridge initial estimates. Our findings support previous research (Yuan and Lin 2007), indicating that ridge initial estimates can help mitigate multicollinearity issues. In addition, the number of variables selected by each approach was quite different but the prediction performance was similar, except for classical methods, which performed slightly worse, probably due to their discrete nature and lack of shrinkage (Hastie et al. 2009). The prediction of classical methods can be improved using post-estimation shrinkage factors (Van Houwelingen and Saurbrei 2013). We compared NNG (R, CVopt) fitted values to those of competing approaches and found that NNG (R, CVopt) and Alasso (L, CVopt) were similar, but NNG (R, CVopt) and lasso were not. This might be explained by the nature of the penalty function as shown in Figure 1, which hardly shrinks variables with large effects.

The results of high-dimensional data showed that the choice of initial estimates influenced the performance of the NNG and adaptive lasso. When ridge and lasso initial estimators were used, the two methods chose larger and simpler models, respectively. Simulation studies in high-dimensional settings found that lasso initial estimators had a higher success rate in terms of sparsity recovery than ridge initial estimators when there were only a few true effects (Zhang, Jeng and Liu 2008). This may be attributed to the tuning parameter of the ridge, which controls the quality of the initial estimates (Yuan and Lin 2007). All the approaches with optimal tuning parameters had slightly different predictions, with ridge performing slightly better and relaxed lasso slightly worse. When the relaxation parameter ($\phi$) of the relaxed lasso is zero, it is equivalent to selecting the variables with the lasso and estimating the coefficients with the unpenalized maximum likelihood method (Meinshausen 2007). When compared to the relaxed lasso, with $\phi \neq 0$, this performs poorly in terms of prediction, possibly due to lack of shrinkage. This might explain why relaxed lasso performed slightly worse in the current study because zero was allowed to be an element of the set of relaxation parameters and was chosen in 20% of the cross-validation folds. The NNG and adaptive lasso predictions were nearly identical, confirming findings by Zhang, Jeng and Liu (2008) that both methods can behave in a similar manner when the same initial estimator is used.

**Conclusions**

In LDD with low correlation between variables, the original NNG with OLS initial estimates selected simple models that were easier to interpret, and predictions were very similar to competing approaches. This seems to suggest that the NNG is suitable for deriving models for both description and prediction. Replacing OLS by ridge initial estimates in data with an HDM helped NNG select simpler models while using lasso initial estimates in HDD helped NNG select simpler models than competing approaches. This indicates that the NNG may be a suitable approach for the analysis of highly correlated and high-dimensional data. In our three examples, we confirmed that the lasso tends to select too many variables, and when it selects simple models, the parameter estimates are often over-shrunken. Some of the results of the adaptive lasso were similar to NNG. Neutral comparison simulation studies advocated by Boulesteix, Wilson and Hapfelmeier (2017) are needed to gain further insight into the advantages and disadvantages of approaches combining variable selection with shrinkage. A paper on the protocol of a simulation study was recently published (Kipruto and Sauerbrei 2022) and the simulation study will be conducted soon.

**Acknowledgements**

The authors gratefully acknowledge Georg Heinze [Medical University of Vienna, Austria] and Shuo Wang [Medical Center-University of Freiburg, Germany] for their helpful comments on an earlier draft and Valentin Seithuemmer [Medical Center-University of Freiburg, Germany], and Sarah Hag-Yahia [Medical Center-University of Freiburg, Germany] for some technical help.

**Conflict of Interest**

The authors have declared no conflict of interest.

# Supporting information

## Web Appendix A
**Dataset**
**Prostate Cancer data**

Data on prostate cancer was sourced from a prostate cancer study (Stamey *et. al.* 1989) that examined factors associated with a raised prostate-specific antigen (PSA). The data was made up of the medical records of 97 male patients who were about to receive radical prostatectomy. The response variable was the logarithm of prostate-specific antigen (**Y**), while the covariate variables (**X**) were eight clinical measures: logarithm of cancer volume (x1), logarithm of prostate weight (x2), seminal vesicle invasion (x3), age (x4), logarithm of benign prostatic hyperplasia amount (x5), logarithm of capsular penetration (x6), Gleason score (x7), and percentage Gleason scores 4 or 5 (x8). This dataset are publicly available on the elements of Statistical Learning book's website (https://hastie.su.domains/ElemStatLearn/datasets/prostate.data) and has previously been used in several earlier publications (Tibshirani 1996; Yuan and Lin 2007) with some covariates logarithm transformed; a form of initial data analysis often carried out before formal statistical analysis (Huebner *et. al.* 2018). In the present paper, the aforementioned transformed data was used.

**Body fat data**

Body fat data (Johnson 1996) contains records of physical and body circumference measurements for 252 men. The outcome of interest was the percentage of body fat (**Y**) with 13 covariates: age (x1), weight (x2), height (x3), and the 10 body circumference measurements: neck (x4), chest (x5), abdomen (x6), hip (x7), thigh (x8), knee (x9), ankle (x10), biceps (x11), forearm (x12), and wrist (x13). The data is available on the multivariable fractional polynomial website (https://mfp.imbi.uni-freiburg.de/book#dataset_tables).

Gene expression data

The pre-processed data set reported in Boulesteix, Guillemot and Sauerbrei (2011) was used, and the data is available on the Institute for Medical Information Processing Biometry and Epidemiology (IBE) website

(https://www.ibe.med.uni-muenchen.de/organisation/mitarbeiter/020_professuren/boulesteix/cvcomplexity/index.html)

# Web Appendix B

**Methods**

**Lasso**

For the classical regression model, the lasso regression estimates are obtained by solving

$$\hat{\beta}^{lasso}(\lambda) = \arg\min_c \frac{1}{2n} \sum_{i=1}^{n} \left( y_i - \sum_{j=1}^{p} \beta_j x_{ij} \right)^2 + \lambda \sum_{j=1}^{p} |\beta_j|, \quad \lambda \geq 0$$

Where the penalty term shrinks the regression coefficients toward zero, while setting some of the coefficients to be exactly equal to zero, thus conducting variable selection (Tibshirani 1996). Unlike the original nonnegative garrote, the lasso can be applied to both low and high-dimensional data. Several authors have studied the properties of the lasso. Zou (2006) showed that the lasso in general is not variable selection consistent unless the design matrix satisfies a strong assumption, the so-called irrepresentable condition (Buehlmann and van de Geer 2013). Zou and Hastie (2005) showed that in high-dimensional data the lasso selects at most n variables, which is undesirable property especially if the true data-generating model consists of more than $n$ covariates. Several extensions of the lasso have been proposed to improve its performance such as the adaptive lasso and relaxed lasso.

**Adaptive lasso**

The adaptive lasso was proposed by Zou (2006) and it modifies the lasso penalty by assigning different weights to different coefficients. This implies that initial estimates are needed to construct adaptive weights. The penalty term imposes severe shrinkage to small coefficients while large coefficients are hardly shrunk (Kipruto and Sauerbrei 2022). For classical linear regression model, the adaptive lasso regression coefficients are obtained by optimizing:

$$\hat{\beta}^{Alasso}(\lambda) = \arg\min_\beta \frac{1}{2n} \sum_{i=1}^{n} \left( y_i - \sum_{j=1}^{p} \beta_j x_{ij} \right)^2 + \lambda \sum_{j=1}^{p} w_j |\beta_j|, \lambda \geq 0$$

where $w_j = 1/\left|\hat{\beta}_j^{init}\right|^\gamma$ is an adaptive weight for the $j^{th}$ variable , $\hat{\beta}_j^{init}$ is an initial estimator, and $\gamma > 0$. Zou (2006) proved that for a fixed number of variables the adaptive lasso has an oracle property in that as $n \to \infty$, the selected set of variables approaches the true set with probability tending to 1. In addition, the estimators are asymptotically normal with the same mean and covariance that they would have by maximum likelihood estimation when the correct submodel is known in advance. Moreover, he showed that the adaptive lasso with $\gamma = 1$ is closely related to the nonnegative garrote and their shrinkage behavior is similar as shown in Figure 1 (bottom-left panel).

**Relaxed lasso**

The relaxed lasso was proposed by Meinshausen (2007) with the aim of reducing the number of noise variables selected by the lasso when cross-validation is used to select the optimal tuning parameter as well as reduce the estimation bias of nonzero coefficients caused by over-shrinkage (Kipruto and Sauerbrei 2022).

The simplified version of the relaxed lasso estimator is given by

$$\hat{\beta}^{relax}(\lambda, \phi) = \phi\hat{\beta}^{lasso}(\lambda) + (1-\phi)\hat{\beta}^{OLS}$$

where $\hat{\beta}^{OLS}$ denotes the OLS estimates obtained by regressing Y on the set of covariates selected by the lasso ($X_{A_\lambda}$), padded with zeros to match the zeros of the lasso solution and $\phi$ is the relaxation parameter that controls the amount of shrinkage of coefficients (Hastie, Tibshirani and Tibshirani 2020). When $\phi = 1$, the relaxed lasso and lasso are identical, and when $\phi < 1$, the amount of shrinkage of coefficients in the selected model is reduced compared to the lasso, as shown in Figure 1 (bottom-right panel). Similarly, when $\phi = 0$, the relaxed lasso estimator is equivalent to the OLS estimators based on the linear model with covariates in $A_\lambda$ (Meinshausen 2007, Kipruto and Sauerbrei 2022).

**Tuning Parameters for lasso, adaptive lasso and relaxed lasso**

The lasso was tuned over 100 values of $\lambda$ as per the default in $cv.glmnet$ function in the glmnet package (Friedman, Hastie and Tibshirani 2010) in R software (R Core Team 2021) and the optimal tuning parameter was selected. Adaptive lasso has two tuning parameters, $\lambda$ and $\gamma$, provided the initial estimates ($\hat{\beta}^{init}$) are given, thus a two-dimensional 10-fold CV was used to find the optimal pair $(\lambda, \gamma)$. We evaluated four values of $\gamma = (0.5, 1.0, 1.5, 2.0)$ and 100 values of $\lambda$. The relaxed lasso has two tuning parameters $\lambda$ and $\phi$. The optimal pair $(\lambda, \phi)$ was obtained via a two-dimensional 10-fold CV where 100 values of $\lambda$ and five values of $\phi = (0, 0.25, 0.5, 0.75, 1)$ were evaluated as per the default in the *cv.glmnet* function. In all procedures, tuning was performed by minimizing mean squared error and deviance in Gaussian and binomial models, respectively.

**Best subset selection**

Best subset selection is one of the classical methods of variable selection that involves identifying subsets of $p$ covariates via an exhaustive search that fits the data well (Miller 2002). The regression estimates of the best-fitting subset of size $k$ are the solution to the constrained optimization problem

$$\hat{\beta}^{subset} = \arg\min_{\beta} \frac{1}{2n} \sum_{i=1}^{n} \left( y_i - \sum_{j=1}^{p} \beta_j \, x_{ij} \right)^2, \text{subject to} \sum_{j=1}^{p} 1(\beta_j \neq 0) \leq k$$

Where $1(\cdot)$ is an indicator function that takes the value of one when a variable is selected and zero otherwise (Kipruto and Sauerbrei 2022). Best subset selection has three disadvantages. First, when the number of variables are large, the computation becomes infeasible. Second, it is unstable (Breiman 1996, Sauerbrei 1999) and lastly, it can lead to the selection of spurious covariates caused by searching over larger sets of models (James *et. al.* 2013, Sauerbrei *et al.* 2020). Cross-validation (CV) and Bayesian information criterion (BIC) were used to select the final model.

## Web Appendix C

**Software used**

All analyses were conducted using R software version 4.1.2 (R Core Team 2021). The lasso, relaxed lasso, and adaptive lasso were implemented using the glmnet package version 4.1-2 due to its computational efficiency (Friedman *et. al.* 2010). The computational details for the adaptive lasso are explained in Zou (2006) section 3.5. Yuan and Lin (2007) showed that the nonnegative garrote solution path can be solved using a modified least-angle regression (LARS) algorithm (Efron *et. al.* 2004) with non-negativity constraints. Thus, we used a custom-made script that uses glmnet to obtain the optimal shrinkage factors for the nonnegative garrote by constraining the lower bound of parameters to zero. Again, glmnet was used to obtain ridge regression estimates that were required while constructing adaptive lasso and nonnegative garrote weights. The best subset selection method was implemented using the leaps package version 3.1 (Lumley 2020), which uses a pure branch-and-bound algorithm (Furnival and Wilson 2000). The algorithm returns a 'best model' of each size (*i.e.* a model with one variable, two variables, to the model with all variables), where 'best model' is a model with the smallest residual sum of squares for each model size (James *et. al.* 2013). BIC and 10-fold CV were used to select the final model. BIC is automatically implemented in the package while CV is not. Therefore, a custom-made script was used to implement 10-fold CV as explained in James *et. al* (2013). Lastly, the pROC package (Robin *et. al.* 2011) was used to compute the area under the receiver operating characteristics.

## Web Appendix D

**Results**

**Prostate cancer**

**Web Table 1.** Prostate data. Comparison of NNG (O, CV) sandwich and bootstrap standard errors. A total of 1000 bootstrap replications was conducted.

|  | Original data | | Fixed $\lambda_{opt}$ | | Re-estimated $\lambda_{opt}$ | | Fixed $\lambda_{1se}$ | | Re-estimated $\lambda_{1se}$ | |
|---|---|---|---|---|---|---|---|---|---|---|
| Covariates | $\hat{\beta}_*$ | SE* | $\hat{\beta}_{**}$ | SE** | $\hat{\beta}_{**}$ | SE** | $\hat{\beta}_{**}$ | SE** | $\hat{\beta}_{**}$ | SE** |
| x1 | 0.642 | **0.091** | 0.639 | **0.101** | 0.644 | **0.095** | 0.665 | 0.116 | 0.648 | 0.108 |
| x2 | 0.211 | **0.077** | 0.184 | **0.116** | 0.242 | **0.103** | 0.083 | 0.097 | 0.113 | 0.102 |
| x5 | 0.209 | **0.088** | 0.203 | **0.114** | 0.277 | **0.108** | 0.101 | 0.107 | 0.134 | 0.115 |
| x3 | 0.000 | 0.000 | -0.017 | 0.038 | -0.106 | 0.087 | 0.000 | 0.000 | -0.007 | 0.023 |
| x4 | 0.000 | 0.000 | 0.050 | 0.077 | 0.115 | 0.098 | 0.009 | 0.032 | 0.029 | 0.059 |
| x6 | 0.000 | 0.000 | -0.003 | 0.021 | -0.085 | 0.119 | 0.000 | 0.000 | -0.003 | 0.020 |
| x7 | 0.000 | 0.000 | 0.004 | 0.020 | 0.024 | 0.075 | 0.000 | 0.005 | 0.002 | 0.014 |
| x8 | 0.000 | 0.000 | 0.033 | 0.062 | 0.097 | 0.118 | 0.009 | 0.036 | 0.018 | 0.049 |

*SE\* and SE\*\* denotes sandwich and bootstrap standard error respectively while $\hat{\beta}_*$ and $\hat{\beta}_{**}$ denotes the NNG regression estimate from the original data and bootstrap sample respectively*

**Web Table 2.** Prostate cancer data. Cross-validation MSE with corresponding standard errors (SE) enclosed in brackets and average number of variables with SE selected in CV for several methods using optimal and 1SE rule tuning parameters.

|  | Optimal | | 1SE rule | |
|---|---|---|---|---|
| Method | MSE | #Variable | MSE | # Variable |
| Full ols | 0.546(0.089) | 8.0(0.000) | - |  |
| Rlasso | 0.569(0.088) | 4.9(0.722) | 0.577(0.073) | 2.4(0.267) |
| Lasso | 0.554(0.086) | 6.4(0.476) | 0.612(0.074) | 3.6(0.267) |
| NNG(O,CV) | 0.582(0.084) | 5.5(0.601) | 0.584(0.069) | 3.2(0.200) |
| NNG(R,CV) | 0.584(0.083) | 5.5(0.601) | 0.579(0.069) | 3.0(0.000) |
| NNG(L,CV) | 0.585(0.083) | 5.3(0.559) | 0.578(0.070) | 3.0(0.000) |
| NNG(O,AIC) | 0.562(0.083) | 5.9(0.180) | - | - |
| NNG(O,BIC) | 0.552(0.079) | 4.5(0.373) | - | - |
| Alasso(O,CV) | 0.577(0.085) | 4.7(0.597) | 0.599(0.069) | 2.7(0.367) |
| Alasso(R,CV) | 0.581(0.085) | 5.0(0.596) | 0.588(0.072) | 2.8(0.200) |
| Alasso(L,CV) | 0.580(0.085) | 4.7(0.597) | 0.591(0.070) | 2.6(0.267) |
| BS(CV) | 0.571(0.090) | 5.1(0.752) | - | - |
| BS(CV+PWSF) | 0.559(0.084) | 5.1(0.752) | - | - |
| BS(BIC) | **0.515**(0.080) | 3.0(0.00) | - | - |
| BS(BIC+PWSF) | 0.516(0.077) | 3.0(0.00) | - | - |

# Body fat

Web Table 3. **Body fat data. Standardized regression coefficients for different methods. Optimal tuning parameters were used in penalized regression methods.**

| Variable | $\hat{\beta}^{OLS}$ | P | NNG(O) | NNG(L) | NNG(R) | Alasso(O) | Alasso(R) | Alasso(L) | Lasso | Rlasso |
|---|---|---|---|---|---|---|---|---|---|---|
| x1 | 8.95 | **0.00** | 8.31 | 7.47 | 7.44 | 8.60 | 8.38 | 7.43 | 7.34 | 7.71 |
| x2 | -1.65 | **0.00** | -1.56 | -1.70 | -1.69 | -1.58 | -1.59 | -1.72 | -1.46 | -1.76 |
| x3 | 0.97 | **0.02** | 0.83 | 1.05 | 1.05 | 0.92 | 0.95 | 1.09 | 0.82 | 0.72 |
| x4 | -0.92 | 0.09 | -0.73 | -0.99 | -0.94 | -0.86 | -0.85 | -1.04 | -0.59 | - |
| x9 | 0.58 | 0.16 | 0.35 | 0.46 | 0.43 | 0.45 | 0.45 | 0.50 | 0.22 | - |
| x7 | 1.02 | 0.17 | 0.57 | 0.54 | 0.57 | 0.91 | 0.76 | 0.58 | 0.32 | - |
| x8 | -1.05 | 0.26 | -0.57 | - | - | -1.00 | -0.63 | - | | - |
| x5 | -0.55 | 0.27 | -0.62 | -0.77 | -0.77 | -0.58 | -0.70 | -0.77 | -0.75 | -0.79 |
| x10 | 0.52 | 0.30 | 0.20 | 0.29 | 0.26 | 0.25 | 0.32 | 0.34 | 0.18 | - |
| x6 | -0.88 | 0.31 | -0.57 | - | - | -0.77 | -0.80 | - | | - |
| x13 | 0.28 | 0.43 | - | - | - | - | - | - | - | - |
| x11 | -0.68 | 0.71 | - | - | - | - | - | - | - | - |
| x12 | 0.02 | 0.97 | - | - | - | - | - | - | - | - |
| # variables | | 13 | 10 | 8 | 8 | 10 | 10 | 8 | 8 | 4 |
| $R^2$ | | 0.74 | 0.74 | 0.74 | 0.74 | 0.74 | 0.74 | 0.74 | 0.73 | 0.73 |
| $Adj.R^2$ | | 0.73 | 0.73 | 0.73 | 0.73 | 0.73 | 0.73 | 0.73 | 0.73 | 0.72 |

Web Table 4. Body fat data. Cross-validation mean-squared errors (MSE) with corresponding standard errors enclosed in brackets and average number of variables with standard errors selected in cross-validation for several methods using optimal and one standard error rule tuning parameters.

| | Optimal $\lambda$ | | 1SE rule $\lambda$ | |
|---|---|---|---|---|
| Method | MSE | # Variable | MSE | # Variable |
| Full ols | 19.324(1.187) | 13.0(0.000) | - | - |
| Rlasso | 19.384(1.253) | 5.4(0.581) | **19.491**(1.378) | 3.4(0.306) |
| Lasso | 19.255(1.242) | 8.9(0.690) | 20.156 (1.378) | 4.0(0.000) |
| NNG(O,CV) | 19.477(1.177) | 8.8(0.772) | 20.384(1.314) | 3.1(0.314) |
| NNG(R,CV) | **19.229**(1.267) | 6.7(0.578) | 20.100(1.235) | 3.8(0.133) |
| NNG(L,CV) | 19.497(1.257) | 7.7(0.633) | 20.086(1.249) | 3.2(0.249) |
| NNG(O,AIC) | 19.395(1.137) | 9.5(0.401) | - | - |
| NNG(O,BIC) | 19.411(1.088) | 5.2(0.291) | - | - |
| Alasso(O,CV) | 19.508(1.215) | 8.3(0.633) | 20.446(1.317) | 2.5(0.167) |
| Alasso(R,CV) | **19.135**(1.146) | 6.2(0.757) | 20.279(1.278) | 3.2(0.249) |
| Alasso(L,CV) | 19.394 (1.197) | 7.2(0.646) | 20.341(1.328) | 2.7(0.260) |
| BS(CV) | 20.017 (1.522) | 9.1(1.178) | - | - |
| BS(CV+P) | 19.676 (1.394) | 9.1(1.178) | - | - |
| BS(BIC) | 19.710 (1.142) | 3.3(0.153) | - | - |
| BS(BIC+P) | 19.685(1.161) | 3.3(0.153) | - | - |

## Gene expression

**Web Table 5**. Gene expression data. Descriptive statistics of the number of variables selected in 10-fold cross-validation using prediction optimal and 1SE rule tuning parameters

|         | Optimal tuning parameters | | | | 1SE tuning parameters | | | |
|---------|------|----|------|------|------|----|------|------|
| Method  | Mean | SD | Min  | Max  | Mean | SD | Min  | Max  |
| Ridge   | 22,283 | 0 | 22,283 | 22,283 | 22,283 | 0 | 22,283 | 22,283 |
| Lasso   | 69   | 27 | 6    | 98   | 10   | 17 | 0    | 53   |
| Rlasso  | 41   | 42 | 1    | 98   | **7** | 16 | 0    | 52   |
| Alasso(R) | 106 | 8  | 91   | 124  | 82   | 9  | 69   | 93   |
| Alasso(L) | 45  | 17 | 3    | 69   | 36   | 18 | 2    | 63   |
| NNG(R)  | 118  | 11 | 100  | 134  | 85   | 11 | 71   | 104  |
| NNG(L)  | **37** | 13 | 4  | 54   | 31   | 12 | 2    | 47   |

# Web Appendix E

## Abbreviations

1SE: one standard error; AIC: Akaike information criterion; Alasso: adaptive lasso; BIC: Bayesian information criterion ; BS: Best subset selection; CV: cross-validation; IBE: Institute for Medical Information Processing Biometry and Epidemiology; LARS: least-angle regression; LOOCV: leave-one-out cross-validation; MSE: mean squared error; NNG: nonnegative garrote; PSA: prostate-specific antigen; Rlasso: relaxed lasso; RSS: residual sum of squares; SE: standard error.